\documentclass[twocolumn]{aastex631}
\usepackage{graphicx}
\usepackage[T1]{fontenc}
\usepackage{amssymb}
\usepackage{amsmath}
\usepackage{CJK}

\newcommand{\revise}[1]{{#1}}

\begin{document}

\title{Earths are not Super-Earths, Saturns are not Jupiters:\\Imprints of pressure-bump planet formation on planetary architectures}

\author[0000-0002-9408-2857]{Wenrui Xu \begin{CJK*}{UTF8}{gbsn}(许文睿)\end{CJK*}}
\affiliation{Center for Computational Astrophysics, Flatiron Institute, New York, USA}
\author[0000-0002-7846-6981]{Songhu Wang}
\affiliation{Department of Astronomy, Indiana University, Bloomington, USA}

\begin{abstract} 
In protoplanetary disks, sufficiently massive planets excite pressure bumps, which can then be preferred locations for forming new planet cores. We discuss how this loop may affect the architecture of multi-planet systems, and compare our predictions with observation. Our main prediction is that low-mass planets and giant planets can each be divided into two subpopulations with different levels of mass uniformity. Low-mass planets that can and cannot reach the pebble isolation mass (the minimum mass required to produce a pressure bump) develop into intra-similar ``Super-Earths'' and more diverse ``Earths'', respectively. Gas giants that do and do not accrete envelope quickly develop into intra-similar ``Jupiters'' and more diverse ``Saturns'', respectively. Super-Earths prefer to form long chains via repeated pressure-bump planet formation, while Jupiter formation is usually terminated at pairs or triplets due to dynamical instability.
These predictions are broadly consistent with observations. In particular, we discover a previously overlooked mass uniformity dichotomy among the observed populations of both low-mass planets (Earths vs. Super-Earths) and gas giants (Saturns vs. Jupiters). For low-mass planets, planets well below the pebble isolation mass ($\lesssim 3M_\oplus$ or $\lesssim 1.5 R_\oplus$ for sun-like stars) show significantly higher intra-system pairwise mass difference than planets around the pebble isolation mass. For gas giants, the period ratios of intra-system pairs show a bimodal distribution, which can be interpreted as two subpopulations with different levels of mass uniformity. These findings suggest that pressure-bump planet formation could be an important ingredient in shaping planetary architectures.
\end{abstract}


\section{Introduction}\label{sec:intro}

Local pressure maxima (pressure bumps) in protoplanetary disks have received increasing attention in recent years, as they are important for both theory and observation of planet formation. Acting as dust traps, pressure bumps are directly related to the rich radial substructures that have been observed in dust continuum observations of protoplanetary disks \citep{dsharpII,dsharpVI,Teague2018}. Pressure bumps (and the resulting rings and gaps) can be a signature of planets in young disks, as planet-disk interaction is a promising (thought not the only) mechanism for generating pressure bumps \citep{Rice2006,dsharpVII,Dong2018}. Moreover, pressure bumps are likely promising venues for forming planet cores. A pressure bump can stop the radial drift of pebbles and increase the pebble surface density, which promotes both planetesimal formation 
and subsequent pebble accretion (e.g., \citealt{Carrera2021,XuBai2022}).
Despite uncertainties on the quantitative details of this process, many studies found that pressure bumps allow massive ($\sim 10M_\oplus$) cores to form over a timescale of $t_{\rm core}\sim 10^4$--$10^5$~yr for pressure bumps located at $r_{\rm bump}\sim3$--15~au \citep{Guilera2020,Morbidelli2020,Chambers2021,Andama2022,Lau2022}. In this case, core formation is effectively instantaneous compared to the disk evolution timescale $t_{\rm disk}\sim$~Myr. Core formation will be slower if the ring is located at a larger radius, and several studies found $t_{\rm core}\sim 10^5$--$10^6$~yr for $r_{\rm bump}\sim10^{2}$~au \citep{Andama2022,Lau2022,Jiang2023}.


Combining existing studies of pressure bump formation by planet-disk interaction and planet formation at pressure bumps, we arrive at an interesting conclusion: \textit{planets and pressure bumps promote the formation of each other, thus they may form a positive feedback loop.} The goal of this study is to discuss the potential effect of this loop on the architecture of multi-planet systems. A major challenge is that this problem involves many different physical ingredients, such as pebble formation by collisional growth, the formation of planetesimals, the dynamical evolution of pebbles and planetesimals, planet-disk (and planetesimal/embryo-disk) interaction and migration. There are still important uncertainties on both the initial conditions of the disk, and the details of some of these processes (see a recent review in \citealt{Drazkowska2023} \S3; also see comments in \citealt{Chambers2021} \S 4.6 and \citealt{Jiang2023} \S 4.1).
It is also technically challenging to self-consistently model all these ingredients in a single calculation (but see \citealt{Chambers2021} and Lau et al. in prep for some recent attempts). In this study we refrain from such comprehensive modeling; instead, we focus on answering a simpler question: Under the simple (though still uncertain) assumption that planet core formation at pressure bumps are rapid ($t_{\rm core}\ll t_{\rm disk}$), can we combine this assumption with existing theoretical knowledge on planet formation and planet-disk interaction to produce some basic predictions that are directly testable on observation?

The remaining of the paper is organized as follows. In Section \ref{sec:superearths} and \ref{sec:giants}, we focus on low-mass planets and giant planets, respectively; in each section we first predict a few trends on the outcome of planet formation that pressure bumps can produce, then compare these predictions against observations,\footnote{\revise{Scripts used for data reduction and plotting are available at \url{https://github.com/wxu26/uniformity_dichotomy}.}} and conclude by discussing whether these observational trends have been reproduced by alternative mechanisms in previous studies. In Section \ref{sec:conclusion} we summarize our results and highlight some future directions.

\begin{figure*}
    \centering
    \includegraphics[width=\textwidth]{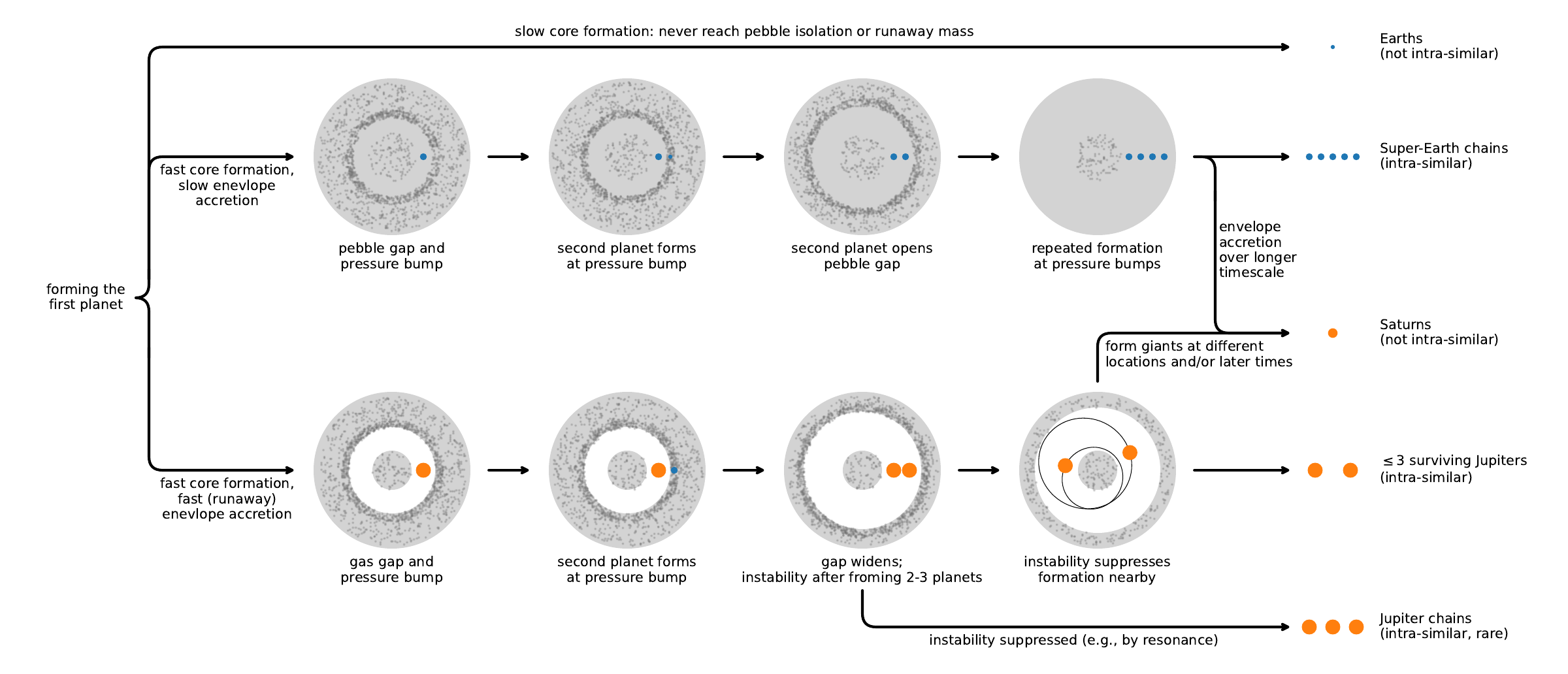}
    \caption{Schematic diagram of various outcomes of planet formation when we assume efficient planet formation at pressure bumps. This diagram summarizes the results from Section \ref{sec:superearths:th} and \ref{sec:giants:th}. Pressure bumps allow planets to form quickly and at very similar time and location, which leads to pairs or chains of planets with highly similar masses. This leads to intra-similar ``Super-Earths'' and ``Jupiters'' whose masses are set by the pebble isolation mass and gas saturation mass, respectively. Meanwhile, planets that form cores slowly (``Earths'') or accrete massive envelopes slowly (``Saturns'') would exhibit less intra-system mass uniformity.}
    \label{fig:diagram}
\end{figure*}

\section{Low-mass planets: intra-similar chains of Super-Earths}\label{sec:superearths}
\subsection{Theoretical predictions}\label{sec:superearths:th}

Existing studies of planet formation at pressure bumps often consider pressure bumps produced by a deep gas gap resembling that carved out by a giant planet \citep{Guilera2020,Morbidelli2020,Andama2022,Lau2022,Jiang2023}. However, pressure bumps can also be produced by planets with lower mass.
The pebble isolation mass, defined as the minimum planet mass for creating a pebble-trapping local pressure maximum in the disk, is primarily set by the disk's aspect ratio $h/r$ \citep{Lambrechts2014,Bitsch2018}
\begin{equation}
    m_{\rm iso} \sim 10 M_\oplus \left(\frac{h/r}{0.04}\right)^3 \left(\frac{M_\star}{M_\odot}\right).\label{eq:m_iso}
\end{equation}
There are also some weak dependence on the slope of the disk's unperturbed pressure profile, the level of viscosity, and the size of the pebble, and a more detailed empirical formula is given in \citet{Bitsch2018}.
At low level of disk turbulence ($\alpha<10^{-3}$; see a review in \citealt{Lesur2022} \S5.2), pressure bump produced by a low-mass planet can already create a significant pile-up of pebbles into a dense ring (cf. \citealt{dsharpVII,Dong2018}), which could help trigger rapid planet formation.
We comment that spontaneously forming at least one planet that reaches $m_{\rm iso}$ well before disk dispersal is quite feasible in terms of the pebble accretion rate. For example, simulations in \citet{Lambrechts2019} suggest that it requires only a modestly high pebble flux of $\sim 200 M_\oplus/{\rm Myr}$.
That requirement might be further relaxed if we consider more advantageous locations for pebble generation and planet formation such as the snowline \citep[cf.][]{Drazkowska2017}.
\revise{One caveat is that reaching $m_{\rm iso}$ would be harder if the planet first encounters the ``flow isolation mass'' \citep{RosenthalMurrayClay2020}. However, the importance of the flow isolation remains uncertain because the ratio between the flow isolation mass and the pebble isolation mass depends on the pebble size and contains an unknown order-unity factor that needs to be calibrated with future simulations.}


Now we consider the outcome of low-mass planet formation provided that a first planet forms sufficiently quickly and reaches the pebble isolation mass ($m_{\rm iso}$) well before disk dispersal.
The effect of reaching the pebble isolation mass (thus creating a pressure bump) is twofold.
First, the pressure bump traps the incoming pebble flux, stalling the accretion of the first planet.\footnote{In general, a planet may produce multiple pressure bumps and multiple rings and gaps when the disk viscosity is low \citep[e.g.,][]{Dong2018,dsharpVII}. However, in most cases the secondary pressure bumps that appear at low viscosity are interior to the main pressure bump \citep[e.g., Fig. 1 of][]{Dong2018}; therefore, the first pebble-trapping pressure bump encountered by the incoming pebble flux remains the same. As long as this holds, the exact ring-gap morphology would not affect our argument.}
Second, the trapped pebbles around the pressure bump promotes the formation of a second planet. Given our assumption of rapid planet formation at pressure bumps, this second planet would soon also reach $m_{\rm iso}$, allowing the above cycle to repeat itself.
We comment that in the absence of further planet formation, the first planet may still gain substantial mass after reaching $m_{\rm iso}$ because the incoming pebbles will only be trapped at the pressure bump for a finite (though relatively long) time before eventually diffusing through the gap and get a chance to be accreted \citep[this is often discussed in the context of gas-gap-opening planets; e.g.,][]{Pinilla2012}. However, this process could be suppressed by rapidly forming a second planet at the pressure bump, since most incoming pebbles are consumed by the second planet before diffusing through the gap.

Eventually, this produces a closely packed chain of Super-Earths, whose masses are approximately the pebble isolation mass $m_{\rm iso}$.
Planets in this chain are formed in close proximity in both space and time. Consider two adjacent Super-Earths with masses $m_1\approx m_2\approx m_{\rm iso}$ forming at semi-major axes $a_1,a_2$. The radial separation at formation $\Delta_{\rm init}=a_2-a_1$ is approximately the distance between the first planet and its pressure bump, which is
\begin{equation}
\Delta_{\rm init} \sim 2.5 h \sim 3.2 R_{\rm H}\text{~~for ~}m\approx m_{\rm iso}. \label{eq:delta_se}
\end{equation}
Here $R_{\rm H}$ is the mutual Hill radius defined as
\begin{equation}
    R_{\rm H} \equiv \frac{a_1+a_2}{2}\left(\frac{m_1+m_2}{3M_\star}\right)^{1/3}.
\end{equation}
We caution that the numbers in Eq. \eqref{eq:delta_se} are based on a single simulation (the purple line in \citealt{Bitsch2018} Fig 1, which corresponds to $m\approx m_{\rm iso}$), and these numbers could depend on disk properties such as the level of turbulence. However, we have not found a more general parametrization of the pressure bump location in the literature; previous parameter surveys \citep[e.g.,][]{dsharpVII} mainly focus on the gap width, which differs from the planet-bump separation by some uncalibrated order-unity factor.
The temporal separation of formation is approximately the timescale of core formation at pressure bump ($t_{\rm core}$), which we assume to be well below the disk evolution timescale ($t_{\rm disk}$). Since the characteristic spatial and temporal scale for the variation of $m_{\rm iso}$ (which is mainly set by $h/r$) are $r$ and $t_{\rm disk}$, respectively, adjacent planets in the chain would have a small relative mass difference of order $\mathcal O(\Delta_{\rm init}/a_1)+\mathcal O(t_{\rm core}/t_{\rm disk})$, \revise{provided that they form in the inner disk ($\lesssim 10^1$~au) where both $h/r$ and $t_{\rm core}/t_{\rm disk}$ (see \S\ref{sec:intro}) are small.}
In summary, pressure bumps allow the formation of Super-Earth chains with a high level of intra-system mass uniformity (Fig. \ref{fig:diagram}).

Another interesting prediction of this formation scenario is that Super-Earth chains are dynamically overpacked at formation. The critical separation below which initially circular planet orbits become dynamically unstable is \citep{Gladman1993, PuWu2015}
\begin{equation}
    \Delta_{\rm stb} \approx \begin{cases}
      3.5 R_{\rm H} & \text{two planets}\\
      10 R_{\rm H} & \text{$\geq 3$ planets}
    \end{cases}.
    \label{eq:delta_stb}
\end{equation}
Therefore, newly formed Super-Earth chains with $\Delta_{\rm init}\sim 3.2R_{\rm H}$ (Eq. \ref{eq:delta_se}) should be dynamically unstable.
However, the amplitude of instability is limited by the eccentricity damping arising from planet-disk interaction; in Appendix \ref{a:damping} we estimate the amplitude of the equilibrium eccentricity $e_{\rm eq}$ when excitation balances damping, and find
\begin{equation}
    e_{\rm eq} \sim0.03 \left(\frac{h/r}{0.04}\right)^4 \left(\frac{M_\star}{1~M_\odot}\right) \left(\frac{\Sigma_{\rm g}}{10^3~{\rm g}~{\rm cm}^{-2}}\right)^{-1} \left(\frac{a}{1~\rm au}\right)^2.\label{eq:e_eq_text}
\end{equation}
Therefore, as long as Super-Earths form early when the gas surface density $\Sigma_{\rm g}$ is still high, the overpacked chain would remain in a low-eccentricity state and avoid scattering or collision. This also allows chain formation to continue, as opposed to being disrupted by highly eccentric planets.
Still, the finite amplitude of the instability allows the orbital spacing to diffuse stochastically; such stochastic diffusion slows down when the planets approach a more stable configuration, and becomes negligible once the planets reach marginally stable separations \citep[][\S 5.1]{Deck2013}.
This provides a mechanism to dynamically relax the planets into a marginally stable state before disk dispersal.
We comment that the above process is not limited to Super-Earths and applies to overpacked low-mass planets in general; we plan to perform a more detailed investigation of this process with numerical simulations in the future.
In summary, we predict that Super-Earth chains are initially overpacked and eventually relax into marginally stable separations. For Super-Earth chains with nearly uniform masses, this also implies nearly uniform separation as the marginally stable separation is proportional to the mutual Hill radii.

\begin{figure*}
    \centering
    \includegraphics[scale=0.66]{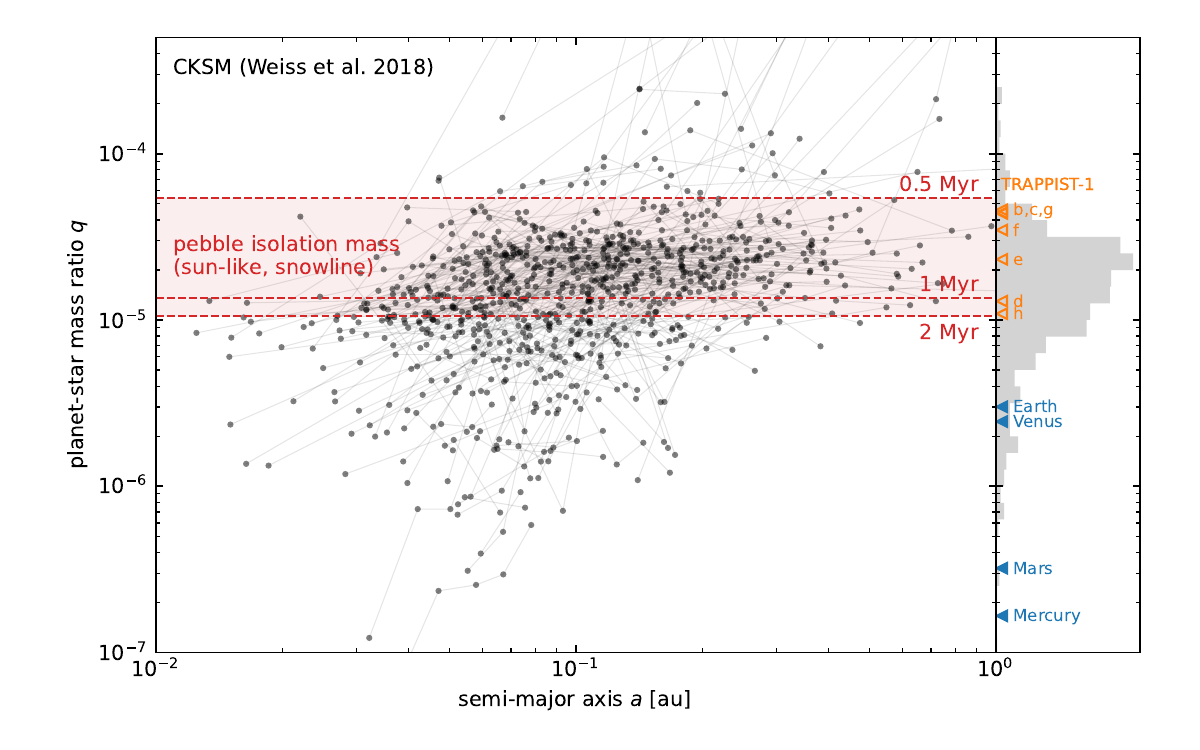}
    \caption{Comparison between the observed planet-star mass ratio $q$ (markers; planets in the same system are also connected by lines) and the pebble isolation mass (red dotted line and red shade). Chain formation at pressure bumps lead to planet masses comparable to the pebble isolation mass at the location and time of formation, which is consistent with the mass scale of Super-Earths.
    The pebble isolation mass is taken from the fiducial model in \citet[Fig. 1]{Bitsch2019} at snowline, with the three red dashed lines corresponding to 0.5, 1, and 2~Myr after disk formation, respectively. (We consider the snowline as opposed to the planet's current location because the snowline is an advantageous location for planet formation, and the planets could migrate after formation.) The observed planets are taken from the California-Kepler Survey multi-planet sample \citep[CKSM;][]{Weiss2018}, and we adopt the mass-radius relation of \citet{ChenKipping2017}. The distribution of planet-star mass ratio of the this sample is shown in the right panel. The peak of this distribution is consistent with the estimated range of pebble isolation mass. We also show the planet-star mass ratio of the TRAPPIST-1 planets (orange markers; \citealt{Agol2021}) and the rocky planets in the solar system (blue markers) for reference. 
    The Earth-mass planets in the low-mass system TRAPPIST-1 resembles Super-Earths around sun-like stars in terms of mass ratio and intra-system uniformity.
    Meanwhile, the rocky planets in the solar system have lower $q$ and exhibit less intra-system uniformity, similar to other low-$q$ exoplanets (cf. Fig. \ref{fig:superearth_uniformity}).
    }
    \label{fig:superearth}
\end{figure*}

\begin{figure}
    \centering
    \includegraphics[scale=0.66]{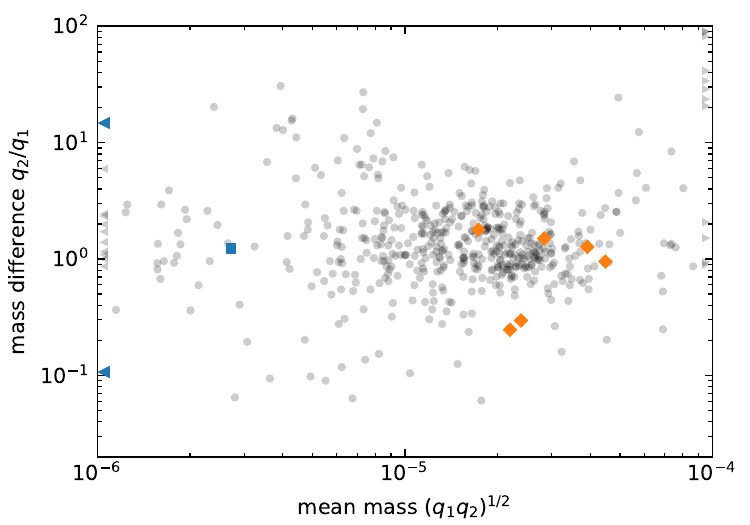}
    \includegraphics[scale=0.66]{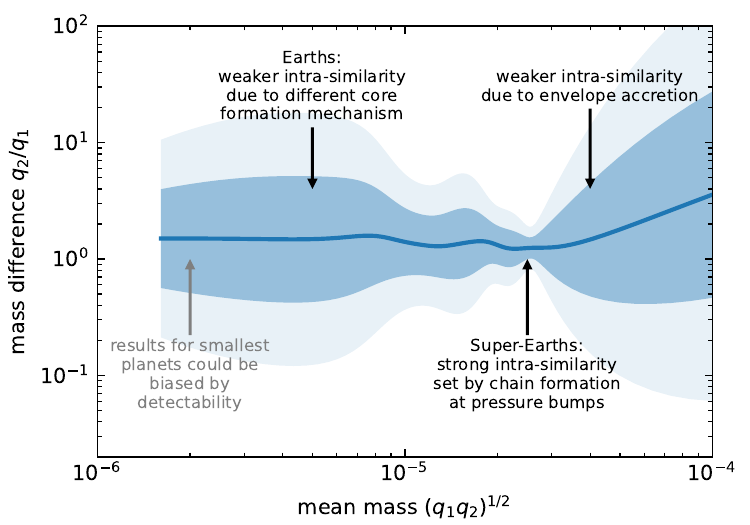}
    \caption{The mass difference between adjacent planet pairs $q_2/q_1$ ($q_{1/2}$ denotes the planet-star mass ratio of the inner/outer planet) versus the mean mass $(q_1q_2)^{1/2}$. Top panel: individual pairs, with marker colors same same as those in Fig. \ref{fig:superearth}. Bottom panel: The distribution of $q_2/q_1$ around given $(q_1q_2)^{1/2}$ for the CKSM population (see method in Appendix \ref{a:q_distribution}).
    We approximate the distribution with a log-normal distribution, with the line showing the (logarithmic) mean and the shaded areas covering $\pm1\sigma$ and $\pm2\sigma$. Our theory predicts that low-mass planets around the pebble isolation mass (Super-Earths) and those well below the pebble isolation mass (``Earths'') are formed via fundamentally different channels, thus the two subpopulations can show different levels of mass uniformity; i.e., ``Earths'' are not Super-Earths. This is in good agreement with the observed distribution. Planets larger than Super-Earths also show less uniformity, which could be due to envelope accretion.
    }
    \label{fig:superearth_uniformity}
\end{figure}

\subsection{Consistency with observation}\label{sec:superearths:obs}

Super-Earths are known to show a ``peas-in-a-pod'' architecture, where Super-Earths in the same system have nearly identical mass and nearly uniform, marginally stable period ratio spacing \citep{Weiss2018, Millholland2017, Wang2017, Goyal2022}. This qualitatively matches our prediction.
Additionally, the characteristic planet-star mass ratio $q$ of Super-Earth systems is indeed comparable to the pebble isolation mass computed from disk models (Fig. \ref{fig:superearth}). 
Here we consider the planet-star mass ratio as opposed to the planet mass because the pebble isolation mass, which is the characteristic mass of Super-Earths formed at pressure bumps, is proportional to the stellar mass (Eq. \ref{eq:m_iso}).
This characteristic mass ratio is also broadly consistent with the chain of Earth-mass planets around the $\sim 0.1 M_\odot$ star TRAPPIST-1, providing a natural explanation for why they show intra-system uniformity similar to Super-Earth chains around sun-like stars despite their much lower masses. We also comment that Jupiter's Galilean moons also have relatively uniform masses with $q=2.5$--$7.8\times 10^{-5}$, but it is less clear whether this can be explained by the pebble isolation mass due to uncertainties of the aspect ratio of Jupiter's disk.

Another key prediction is that chain formation at pressure bumps is only available to planets that reach the pebble isolation mass, and if it is 
the dominant source of intra-system uniformity of Super-Earths, then planets with mass well below the pebble isolation mass (which we dub ``Earths'' in this paper) could show much less intra-system mass uniformity. In other words, \textit{Earths are not Super-Earths.} This uniformity dichotomy is visible in Fig. \ref{fig:superearth_uniformity}, which shows an apparent decrease in mass uniformity for low-mass planets with $q\lesssim 10^{-5}$ (for a sun-like star, this corresponds to planet mass $\lesssim 3M_\oplus$ and planet radius $\lesssim 1.5 R_\oplus$). In statistical terms, this dichotomy of mass uniformity can be phrased as a heteroscedasticity of pairwise mass difference; i.e. the scatter of pairwise mass difference is not constant but varies as a function of the planet mass. We test the statistical significance of this heteroscedasticity in Appendix \ref{a:heteroscedasticity}.
As a side note, the lack of uniformity of the solar system's rocky planets is not surprising; they have mass ratios ranging from $q=1.7\times 10^{-7}$ for Mercury to $q=3\times 10^{-6}$ for Earth, and such order-magnitude difference is comparable to the level of mass difference of exoplanets with $q\lesssim 10^{-5}$.

Chain formation at pressure bumps may also contribute to the inter-system variation of Super-Earth properties. Only a small fraction ($\sim 20\%$) of the inter-system variation of Kepler multi-planet systems can be explained by stellar mass and metallicity \citep{Millholland2017}, and the origin of the rest remains uncertain. One possible source of additional inter-system variation is the inherent randomness of when the chain formation begins. To spontaneously form a planet reaching pebble isolation mass (thus triggering chain formation) at relatively low pebble flux, it may require a lucky collision between two small embryos to bypass the bottleneck of pebble accretion (cf. numerical experiments in \citealt{Lambrechts2019}). Given the relatively large number of small embryos in the disk, the chance of having such collision could be high but exactly when it happens could be quite random. This randomness in the epoch of triggering chain formation affects the pebble isolation mass, which can change by a factor of a few during disk dispersal (see the red shade in Fig. \ref{fig:superearth}; also see Fig. 1 of \citealt{Bitsch2019}).

Finally, chain formation at pressure bumps are favorably efficient. Without the pressure bump (or other substructures), the efficiency of converting pebbles to planets are often limited by the inward drift of pebbles. But with the pressure bump suppressing the inward drift, pebbles can be converted to planets at order-unity efficiency once this chain formation process begins. Such high efficiency is favorable, given observational evidences that the dust mass in Class II disks are comparable to the solid mass budget of planets \citep{Najita2014,Manara2018,Dai2020,Lu2020,Mulders2021}.\footnote{One caveat is that younger disks might be more massive, and that relaxes the requirement on planet formation efficiency if planet/planetesimal formation begins early \citep{Xu2022, XuArmitage2023, Xu2023}}

\subsection{Comparison with previous studies}\label{sec:superearths:com}

Direct formation of Super-Earth chains at pressure bumps is only one of many possible explanations to intra-system mass uniformity. In general, any process that sets a characteristic planet mass scale may introduce such uniformity; for Super-Earths with small semi-major axes, a good candidate for such mass scale is the Goldreich mass \citep{Goldreich2004}, which sets the characteristic mass after giant impacts. (See reviews of some other relevant mass scales in \citealt{Weiss2022} \S6 and \citealt{Emsenhuber2023} \S5.) Another example is \citet{BatyginMorbidelli2023}; they hypothesize planet formation at a largely stationary ring of planetesimals, and the planet masses are set by either isolation or inward migration, whichever happens first. \revise{Note that here ``isolation'' refers to he depletion of planetesimals from the local feeding zone of the embryo; this is different from pebble isolation, and the resulting mass depends on the planetesimal surface density.} In both cases, planets attain similar masses as long as they form quickly (compared to the disk evolution timescale). Finally, minimizing the energy of the system under constraints of conserving total mass, total angular momentum, and planet separation also leads to nearly uniform masses \citep{Adams2019}. Many of these studies also include theories and/or simulations showing intra-similar, marginally stable orbital spacing; generally, marginal stability might be interpreted as a natural outcome when mechanisms that reduce instability (e.g., eccentricity damping by the disk, tidal dissipation, removal or merger of planets or embryos) and mechanisms that increase instability (e.g., convergent migration, overpacked initial configuration) coexist.

However, there is one fundamental difference between pressure-bump planet formation and other explanations for Super-Earth architecture. Pressure-bump planet formation is so far the only mechanism that naturally predicts \revise{not only the mass uniformity of Super-Earths but also} the uniformity dichotomy between Earths and Super-Earths (Fig. \ref{fig:superearth_uniformity}), because the mechanism for generating mass uniformity in other theories continue to operate below the pebble isolation mass.
This dichotomy also does not seem to be reproduced by existing population syntheses, which do not include pressure-bump planet formation. For example, in \citet[][Fig. 6, center and right panels]{Mishra2021} the pairwise mass ratio between short-period planets ($P<640$~d) does not increase visibly at low masses, regardless of whether observational biases have been factored in.
Future studies with quantitative, apple-to-apple comparisons between planet formation models with and without pressure-bump planet formation will determine whether it improves the agreement with the observed trends.

\section{Gas giants: unstable Jupiter twins and diverse Saturns}\label{sec:giants}
\subsection{Theoretical predictions}\label{sec:giants:th}

Giant planets are generally massive enough to form pressure bumps and trigger subsequent planet formation. However, the outcome of planet formation, including the level of mass uniformity and dynamical stability, depends on the timescale of envelope accretion.
In the discussion below, we discuss gas giants that undergo fast and slow envelope accretion separately; for simplicity, we dub these two classes of giant planets ``Jupiters'' and ``Saturns'' respectively.\footnote{The naming only reflects that within the same system, a fast-accreting giant likely attains higher mass than a slow-accreting giant; it does not imply that Jupiter and Saturn in the solar system need to be representative members of these classes.}
More precisely, if a gas giant can accrete a massive envelope and open a deep gas gap at a timescale much shorter than the disk evolution timescale, we categorize it as a Jupiter; otherwise we categorize it as a Saturn.
Physically, whether the disk prefers to form Saturns or Jupiters largely depends on the comparison between the pebble isolation mass (which gives the characteristic mass of the core) and the mass threshold for runaway accretion. When the former is higher, Jupiter formation is preferred because core formation can be directly followed by runaway envelope accretion; when the latter is higher, Saturn formation is preferred because the initial phase of envelope accretion would be slow and cooling-limited.

First consider the regime where disk properties favor Jupiter formation. As we discussed in Section \ref{sec:superearths}, mass uniformity can arise when planet masses are set by some approximately constant characteristic mass, such as the pebble isolation mass. For Jupiters, the reduction of accretion after gap-opening causes the planet to saturate at a final mass $m_{\rm final}$ close to the gap-opening mass (\citealt{GinzburgChiang2019}; hereafter GC19). Since the timescale of core formation and envelope accretion are both $\ll$ disk evolution timescale (given our assumption of rapid core formation at pressure bump and our definition of Jupiters), the mass difference of adjacent Jupiters is mainly due to the radial (but not temporal) dependence of $m_{\rm final}$, which scales as $a^{0.3-0.75}$ (GC19 \S3.2.4).
Now consider the separation between adjacent Jupiters at formation. This is given by the separation between a Jupiter and the pressure bump it excites. For the discussion in this section, we stick to the fiducial case of $1M_J$ planets around a $1M_\odot$ star with semi-major axis separation $a_2/a_1 = 1.5$. Here this 1.5 factor comes from numerical simulations in \citet{Pinilla2012}. This gives a $m_{\rm final}$ difference of $13\%$--$36\%$.

This estimate requires a few comments and caveats. Applying the mass scaling in GC19 effectively assumes that both planets form in an unperturbed disk. This is a reasonable assumption, since the planet's final mass is mainly set by the gas density, which is only slightly perturbed at the pressure bump \citep[e.g.,][]{Pinilla2012} because the local pressure maximum lies outside of the gap. Still, this result is only a qualitative demonstration that the separation between the planet and its pressure bump does not degrade the level of mass uniformity too much. It cannot serve as an accurate estimate of the mass difference mainly because this simple estimate does not capture the full complexity of the accretion after gap-opening, where Jupiter pairs likely reside in a wide, shared gap. In the model of GC19 (see their Fig. 1), post-gap-opening accretion contributes $\approx 25\%$ of the planet mass when the disk viscosity is $\alpha\lesssim 2\times 10^{-4}$.

Another important feature of Jupiters is that Jupiter chain formation will likely be terminated by dynamical instability. There are a few different possibilities in how this process takes place exactly, but one robust prediction is that $\leq3$ Jupiters survive in most cases.
With our fiducial semi-major axis separation of 1.5 for $1M_J$ planets, the separation is $\sim 4.6 R_{\rm H}$; comparing this with the stability threshold (Eq. \ref{eq:delta_stb}) suggests that instability will develop once the chain of Jupiters extend to three planets.
Meanwhile, eccentricity damping by the disk can no longer protect Jupiters due to the reduction in surface density within the gap (Eq. \ref{eq:e_eq_text}; note that $e_{\rm eq}\propto \Sigma_{\rm g}^{-1}$).
Another possibility is that a Jupiter pair may become unstable before the third Jupiter is formed, as gap-sharing planets have the tendency to migrate convergently.
This is because the gap appears asymmetric for each planet, and the inner/outer planet see additional gap width (excited by the other planet) in the outer/inner direction; the migration torque from the outer/inner disk is thus reduced compared to the single-planet case. Since the migration torque of the inner/outer spiral excited the planet drives outward/inward migration, this mechanism tends to drive the planets toward each other.
Depending on the exact rates of Jupiter formation and convergent migration, the separation may and may not drop below the two-body stability limit (this requires only $\sim 10\%$ change in semi-major axis ratio) before the formation of a third Jupiter.

In summary, Jupiters tend to become unstable once they become pairs or triplets, and due to the low gap density, this instability (unlike the Super-Earth case) allows significant eccentricity excitation (Fig. \ref{fig:diagram}).
In principle, a wide range of outcomes would be possible from this instability, including removing one of the planets or significantly altering their orbits and/or increasing eccentricity and inclination. While quantitative discussion of the distribution of different outcomes has to be left for future works, one could at least qualitatively conclude that this dynamical instability would suppress planet core formation nearby (spatially and temporally) by stirring up large bodies (pebbles and planetesimals), unless the planets are lucky enough to enter a more stable state (e.g., resonance) before or shortly after they become unstable. We comment that once Jupiters become unstable, the timescale for developing eccentricity (an rough estimate of which being $t_{\rm MMR}$ in Appendix \ref{a:damping}) is generally lower than the timescale for subsequent core formation and envelope accretion; for a Jupiter-like planet, the former is $\sim 10^3{\rm yr}$ while the latter is $\gtrsim 10^4{\rm yr}$ (see \S \ref{sec:intro}).
Thus, unlike the Super-Earths that form in long, low-eccentricity chains, Jupiters mostly form in unstable pairs and triplets; and the instability may remove some of the Jupiters via scattering.

Saturns, which we define as giant planets that undergo slower envelope accretion, differ from Jupiters in several ways. The most important difference is that Saturns have weaker intra-system mass uniformity than Jupiters. 
Slow envelope accretion means that Saturns either reaches the gas-gap-opening mass over a timescale comparable to the disk evolution timescale, or never opens a gap.
In the former case, the characteristic final mass of the planet is still set by gap-opening (GC19).
Consider the epoch when a planet reaches the gap-opening mass $t_{\rm gap}$. The final mass of the planet depends on $t_{\rm gap}$ because it is mainly set by disk properties at gap-opening. Compared to Jupiters, gap-opening Satruns would have a larger scatter of $t_{\rm gap}$ due to slow envelope accretion, and that reduces mass uniformity.
In the latter case, we also expect low level of mass uniformity because there is no longer a characteristic mass scale that the planets converge towards.

Another difference is that while a system usually contains $\leq 3$ surviving Jupiters due to dynamical instability, there can be more Saturns. For example, Super-Earth chains can develop into Saturn chains via slow envelope accretion. Note that this mechanism of forming multiple giants is more available for slow-accreting Saturns than fast-accreting Jupiters because Jupiters tend to be unstable once they reside in deep gaps, as discussed earlier in this Section. Meanwhile, Saturns can often (though not necessarily always) avoid instability for two reasons: First, they either never reach the gap-opening mass or open gaps at later times, allowing more eccentricity damping from the gas disk; second, the slow accretion may also allow more time for the planets to dynamically relax into a less unstable state (see \S\ref{sec:superearths:th}).

We also comment that multiple types of planets may occur in the same system. In particular, while the dynamical instability of Jupiters suppresses nearby planet formation, it is still possible to form Saturns or low-mass planets in a different part of the disk or at a later time, after the instability of the Jupiter pair quiets down.

\begin{figure}
    \centering
    \includegraphics[scale=0.66]{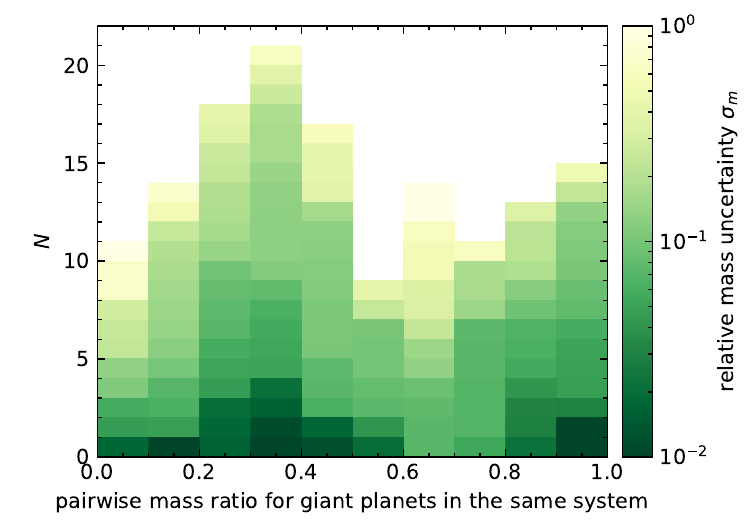}
    \includegraphics[scale=0.66]{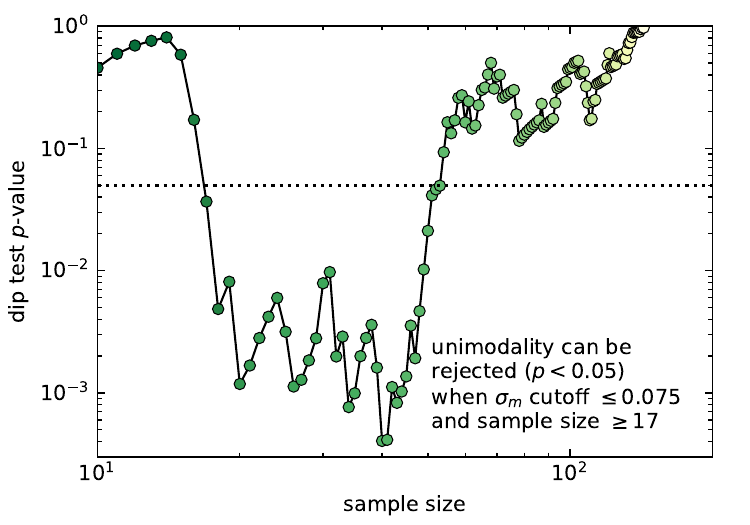}
    \caption{Top panel: the mass ratio distribution of observed intra-system giant planet ($m>100M_\oplus$) pairs. Data is taken from the NASA Exoplanet Archive. Color marks the relative mass uncertainty $\sigma_m$ (Eq. \ref{eq:sigma_m}) of each pair. The mass ratio distribution appears bimodal, which could be a result of distinct formation pathways (``Jupiters'' and ``Saturns''). Bottom panel: examining the significance of \revise{multimodality} with Hartigan's dip test. We perform the test on a range of subsamples; each subsample (each marker in figure) contains all pairs with $\sigma_m\leq$ a certain cutoff value (color coded with same color scale as the top panel). At sufficiently small $\sigma_m$ cutoff (so that \revise{$\sigma_m$ does not diminish the amplitude of bimodality})
    and sufficiently large sample size, the \revise{multimodality} is statistically significant.
    }
    \label{fig:giants_two_pop}
\end{figure}

\subsection{Consistency with observation}\label{sec:giants:obs}

The physical picture discussed above leads to two main predictions. First, giant planets can be divided into two populations: Jupiters, which should exhibit intra-system uniformity (when a system contains more than one of them) but occurs at a relatively low rate due to instability, and Saturns, which show less intra-system uniformity but might be more common. As a result, if we consider the intra-system mass ratio between giant pairs, its distribution would have two distinct components: Jupiter-Jupiter pairs would give a mass ratio distribution centered close to unity, whereas Jupiter-Saturn and Saturn-Saturn pairs would give a broader distribution which does not have to center around unity. In other words, the mass ratio distribution of intra-system giant pairs is likely bimodal. The top panel of Fig. \ref{fig:giants_two_pop} shows that this bimodality does seem to exist among observed giants. We comment that the observed population of giants planets is heavily affected by
\revise{detection biases including the preference for more massive, shorter-period planets and the diversity of detection methods and surveys.} However, the bimodality is likely not an artifact as there is no known mechanism for detection bias to create the relatively narrow dip separating the two modes.

Detecting (and visualizing) the bimodality in Fig. \ref{fig:giants_two_pop} requires some effort because the dip between the two modes can be smeared out if the sample contains too many pairs with large mass uncertainty. We use data in the \citeauthor{exoarchive}, which often contains multiple references for the same planet. While the archive flags one default reference for each planet, this choice is not necessarily the one with the most accurate mass estimate. Additionally, the default references of different planets in the same system may be different, and differences among those references in data reduction and assumed stellar properties introduce a systematic error. To avoid these issues, we reselect the references for each planet pair as follows. First, we check whether the pair has been modeled together in at least one study; if not, the pair is discarded. Then, if there are multiple studies modeling the same pair, we select the study with the lowest relative mass uncertainty
\begin{equation}
    \sigma_m = \sqrt{\sigma_1^2 + \sigma_2^2}.\label{eq:sigma_m}
\end{equation}
Here $\sigma_{1,2}$ is the relative mass uncertainty for each planet. We discard references that only have upper or lower mass limits or do not include mass uncertainty estimates.
Even with this reselection which reduces $\sigma_m$, the sample still contains many pairs with large mass uncertainties. In Fig. \ref{fig:giants_two_pop} top panel we color-code each pair using $\sigma_m$ and choose a colormap that visually emphasizes pairs with more accurate masses.
2We also perform a more quantitative investigation on the statistical significance of \revise{multimodality} using the dip test of \citet{hartigan1985dip}. The $p$-value of this test estimates the probability that the data is drawn from the the best-fitting unimodal distribution. We select subsamples from our data with different $\sigma_m$ cutoff and test each subset in the bottom panel of Fig. \ref{fig:giants_two_pop}. As expected, there is a trade-off between uncertainty and sample size: when the $\sigma_m$ cutoff is too small, the sample size would be insufficient; when the $\sigma_m$ cutoff is too large, \revise{it reduces the contrast between the peaks and the dip in the distribution of observed mass ratio, which is is the distribution of true mass ratio convolved with observational uncertainty (with width $\sim\sigma_m$).}
Still, provided that we choose the cutoff appropriately, a wide range of subsamples show statistically significant \revise{multimodality}.
Finally, to make sure that the bimodality is not an artifact due to polling different studies, we performed the same analysis on a recent RV survey \citep{Rosenthal2021} and found similar results.

\begin{figure*}
    \centering
    \includegraphics[scale=.6]{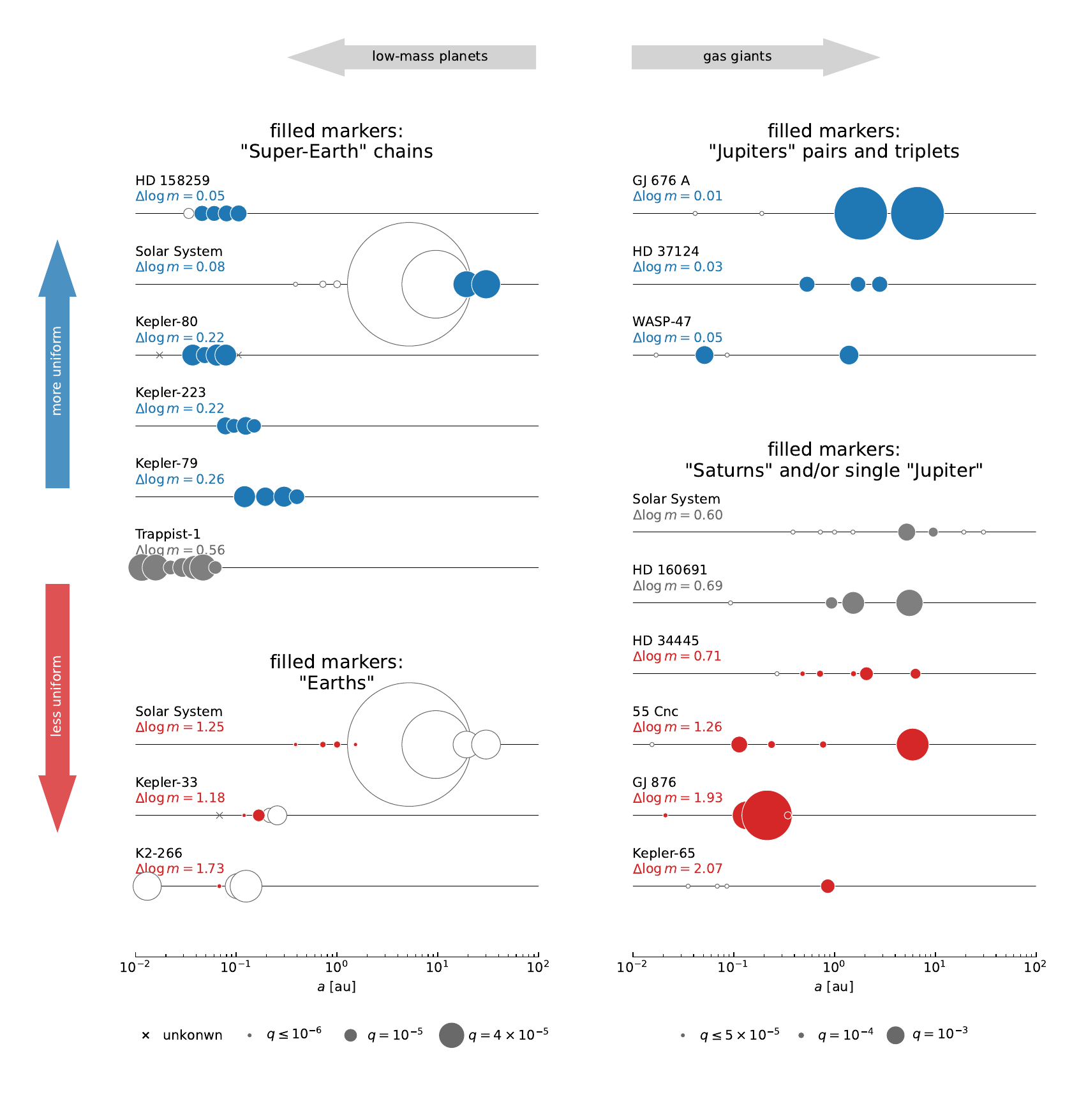}
    \caption{Examples of different classes of planets our theory predicts. This diagram is for illustration only, and the samples shown here may not be statistically representative of the whole planet population. All systems are chosen from a catalog of known systems with $\geq4$ planets compiled in \citet{Mishra2023}, with the exception of HD 37124, which serves as an example of Jupiter triplet. Generally a system can contain multiple classes of planets; we highlight the planets that belong to the relevant class with filled markers and other planets are marked by empty markers. For simplicity, we use the planet-star mass ratio $q$ to (roughly) categorize the planets, with Earths having $q<10^{-5}$, Super-Earths having $q$ between $10^{-5}$ and $q_{\rm Neptune}=5.15\times 10^{-5}$, and Saturns and Jupiters having $q>q_{\rm Neptune}$. We separate the two classes on the right panel by the level of mass uniformity. We list the mass variation $\Delta \log m$ (standard deviation of log planet mass) among the highlighted planets for each system. When there is only one highlighted planet, we compute $\Delta \log m$ using the masses of the highlighted planet and its adjacent planets. The highlight color follows $\Delta \log m$, with blue for $\Delta\log m<\log 1.2$, gray for $
    \log 1.2\leq\Delta\log m\leq \log 2$, and red for $\Delta\log m > \log 2$.  To visualize the level of mass uniformity, marker sizes are chosen to be $\propto q^{1/2}$ ($q$ is the planet-to-star mass ratio) so that the area of the marker is linearly proportional to the planet mass. We use different marker size normalization for low-mass planets (left panel) and gas giants (right panel). We note in passing that Uranus and Neptune are tentatively labeled as Super-Earths here as their masses are very similar to each other and are broadly consistent with the pebble isolation mass. The composition of Uranus and Neptune (which might contain more rock than ice; cf. \citealt{Helled2020,Teanby2020}) could also be similar to that of water-rich Super-Earths \citep[e.g.,][]{Luque2022}.}
    \label{fig:giants_architecture}
\end{figure*}

Another prediction is that the architecture of the systems containing Jupiters can be diverse due to potential scattering events caused by the instability of Jupiter pairs. This might explain the inverted mass-ordering in some systems where one or two smaller planets are found between two larger giants, such as WASP-47 and 55 Cnc (Fig. \ref{fig:giants_architecture}). It may also be helpful for producing the population of hot and warm Jupiters \citep{Wu2023}, and some of the wide-orbit Jupiters observed by direct imaging.

\subsection{Comparison with previous studies}\label{sec:giants:com}

The overabundance of giant planet pairs with nearly equal masses, which we interpret as the signature of a intra-similar Jupiter pairs, is a new discovery. It is reasonable that this trend has been overlooked in existing observational studies, since most surviving giant planets are Saturns and single Jupiters which do not show this intra-similarity.

The existence of intra-similar Jupiter pairs and the diverse architecture that can be produced by their instability does not seem to be recovered by existing population syntheses (which do not include the effect of pressure bumps). For example, in \citet{Mishra2023}, it seems uncommon for synthesized systems (their Fig. 6 right panel) to have giant planets in very similar masses (mass ratio $\gtrsim 0.8$, corresponding to the peak in Fig. \ref{fig:giants_two_pop}). Their synthesized systems also do not show any significant mass inversion (large-small-large configuration, as opposed to the more typical monotonic or small-large-small configuration), which has been observed in a few systems and might be explained by the instability of Jupiter pairs.
This may suggest that Jupiter pair formation due to the pressure bump could be an important ingredient for setting the architecture of systems containing giant planets.

\section{Conclusion and outlook}\label{sec:conclusion}

In this paper we make basic predictions on the outcome of planet formation under the assumption that new planet cores form rapidly at planet-induced pressure bumps, and we test these predictions on observed planet populations.

Fundamentally, pressure-bump aided planet formation provides a mechanism to form two or multiple planets at approximately the same time and location. This causes them to form in nearly identical environments and attain nearly identical masses.
Depending on whether the first planet core grows sufficiently quickly to reach the pebble isolation mass and whether envelope accretion is sufficiently rapid to reach the gap-opening mass, the outcome of planet formation can be divided into for different classes (Sections \ref{sec:superearths:th} and \ref{sec:giants:th}):
\begin{itemize}
    \item ``Earths'': small planets that never reach the pebble isolation mass before disk dispersal. The slow growth makes them exhibit less intra-system mass uniformity than Super-Earths.
    \item ``Super-Earths'': planets whose cores reach the pebble isolation mass but do not accrete much envelope (or undergo envelope removal post-formation). They trigger repeated formation at pressure bumps, leading to chains of planets with $m\approx m_{\rm iso}$ (Eq. \ref{eq:m_iso}). This produces a high degree of intra-system mass uniformity.
    \item ``Saturns'': planets that accrete massive envelopes, but do so slowly. (This include planets that initially form as Super-Earth chains and then slowly accrete massive envelopes.) Their core formation may and may not be aided by pressure bumps, but the slow mass assembly reduces mass uniformity.
    \item ``Jupiters'': planets that accrete envelopes quickly (e.g., when the pebble isolation mass is above the runaway mass). Their growth is limited by gap-opening, and rapid formation and accretion at pressure bumps produce stronger mass uniformity than Saturns. Close separation and low gap density results in dynamical instability, which terminates formation at 2--3 Jupiters and may remove some of them via scattering.
\end{itemize}
We note that the names of these classes are chosen according to the typical planet mass in each class (for sun-like stars), but our definitions do not exactly align with what these names typically imply.
These formation pathways and possible observed examples of each class of planets are summarized in Figs. \ref{fig:diagram} and \ref{fig:giants_architecture}. Note that one planetary system could contain more than one of these classes of planets.

We find several observational evidences in support of these predictions (Sections \ref{sec:superearths:obs} and \ref{sec:giants:obs}). The characteristic planet-to-star mass ratio of observed Super-Earths is in good agreement with the estimated pebble isolation mass around the snowline (Fig. \ref{fig:superearth}). The intra-similarity of Super-Earths do not extend to planets with lower masses (Fig. \ref{fig:superearth_uniformity}), causing a dichotomy between Earths ($q\lesssim 10^{-5}$) and Super-Earths ($q\sim$ few $10^{-5}$). Similarly,
the mass ratio distribution of intra-system giant planet pairs exhibits bimodality,
which could be due to the dichotomy between intra-similar Jupiter-Jupiter pairs and more diverse Saturn-Saturn and Saturn-Jupiter pairs (Fig. \ref{fig:giants_two_pop}). Unstable Jupiter pairs may also help producing the diverse observed orbital architectures, including the population of hot and warm Jupiters and occasional inverted mass ordering. In particular, the ``uniformity dichotimy'' -- that low-mass planets and giants can each be divided into two distinct sub-populations with different level of intra-similarity (Earths vs. Super-Earths, Saturns vs. Jupiters) -- is a new discovery and have not been predicted by existing theoretical studies (cf. Sections \ref{sec:superearths:com} and \ref{sec:giants:com}).

A major limitation of the current study is that the predictions we make are only qualitative, and some of the arguments behind our predictions are yet to be tested by more detailed calculations or simulations. As a result, it would be premature to conclude to what extend the observed architecture of multi-planet systems is shaped by planet formation at pressure bumps. Rather, the main contribution of the current study is to point at a few interesting directions that may help us answer this problem.
These include a closer inspection on the exact location of the pressure bump; investigations of the dynamical stability and long-term orbital evolution of dynamically overpacked planet pairs or chains formed via pressure bumps; and quantitative, apple-to-apple comparison between population synthesis models with and without including the effect of pressure bumps.
Additionally, regardless of whether it is related to pressure bumps, the newly discovered uniformity dichotomy is a new piece of observational constraint which could help testing planet formation theories and population synthesis models.

\section*{Acknowledgments}
We thank the anonymous referee for providing constructive and insightful comments.
W.X. thanks Jiayin Dong, Linn Eirksson, and Philip Armitage for insightful discussions.
S.W. gratefully acknowledges the generous support from the Heising-Simons Foundation, including support from Grant \#2023-4050.
This research has made use of the NASA Exoplanet Archive, which is operated by the California Institute of Technology, under contract with the National Aeronautics and Space Administration under the Exoplanet Exploration Program.

\appendix

\section{Timescale of instability and eccentricity damping}\label{a:damping}
In this appendix we estimate whether the disk can damp the dynamical instability of planets formed at pressure bumps.

The disk suppresses instability by damping planet eccentricity. To lowest order in eccentricity and when the underlying disk surface density profile remains relatively smooth (i.e., no gap opening), the damping timescale is (\citealt{TanakaWard2004}; also see \citealt{CresswellNelson2008})
\begin{equation}
    t_e = 0.78^{-1}q^{-1}\left(\frac{h}{r}\right)^{4}\frac{M_\star}{\Sigma r^2}\Omega^{-1}.\label{eq:t_e}
\end{equation}
This gives
\begin{equation}
    {\rm d}e/{\rm d}t = -e t_e^{-1}.
\end{equation}
Instabilities may also excite inclination, and the damping timescale of inclination is the same as $t_e$ up to an order-unity factor.

To determine whether this damping is enough to suppress instability and avoid close encouters, we need to also estimate the rate of eccentricity excitation. A main mechanism for destablizing close planet pairs is resonance overlap \citep{Deck2013}: when the planets are sufficiently massive and the separation sufficiently small, nearby first-order mean motion resonances (MMRs) can overlap, causing chaos in orbital evolution. The amplitude of ${\rm d}e/{\rm d}t$ of this mechanism can then be estimated by ${\rm d}e/{\rm d}t$ of first-order MMR. First-order MMRs have a characteristic timescale $t_{\rm MMR}$ and eccentricity scale $e_{\rm MMR}$ approximately given by \citep[cf.][]{Goldreich2014}
\begin{align}
    &t_{\rm MMR} \sim \Omega^{-1} q^{-2/3},\\
    &e_{\rm MMR} \sim q^{1/3}.
\end{align}
Here $t_{\rm MMR}$ and $e_{\rm MMR}$ are defined based on the Hamiltonian of the system near a first-order MMR. The separatrix and fixed points of the Hamiltonian \citep[cf.][Fig. 1]{Deck2013} have eccentricities $\sim e_{\rm MMR}$, and the timescale for librating around the fixed point and moving around the spearatrix is $\sim t_{\rm MMR}$. This gives
\begin{equation}
    {\rm d}e/{\rm d}t \sim e_{\rm MMR} t_{\rm MMR}^{-1}.
\end{equation}

Balancing excitation and damping gives a characteristic scale of eccentricity,
\begin{equation}
    e_{\rm eq} \equiv \frac{e_{\rm MMR} t_{\rm MMR}^{-1}}{t_e^{-1}} \sim 0.03 \left(\frac{h/r}{0.04}\right)^4 \left(\frac{M_\star}{1~M_\odot}\right) \left(\frac{\Sigma}{10^3~{\rm g}~{\rm cm}^{-2}}\right)^{-1} \left(\frac{a}{1~\rm au}\right)^2.\label{eq:e_eq}
\end{equation}
This eccentricity is small (for the fiducial values of parameters adopted in the last step) and independent of the planet mass. One caveat here is that the surface density profile can differ significantly across different disk models, especially at small radii ($\lesssim$~few au) where observational constraints are limited.

For gap-opening giant planets, the eccentricity damping rate in Eq. \ref{eq:t_e} needs some modification. A simple estimate is to replace the (unperturbed) disk surface density $\Sigma$ with the surface density inside the gap $\Sigma_{\rm gap}$. This is a reasonable choice, since the damping comes from the excitation of eccentricity waves \citep{TanakaWard2004}, and most such excitation happens close to the planet.\footnote{One caveat is that this has only been tested down to $\Sigma_{\rm gap}/\Sigma = 0.3$ \citep[cf.][]{Kanagawa2018,Pichierri2023}; when $\Sigma_{\rm gap}/\Sigma\ll 1$, replacing $\Sigma$ with $\Sigma_{\rm gap}$ for estimating $t_e$ may become less accurate.} Since the gap depth is probably already several orders of magnitude by the time runaway growth halts (see an example in \citealt{GinzburgChiang2019}), the eccentricity damping will likely become insufficient for suppressing instability (i.e., maintaining $e_{\rm eq}\ll 1$) when the second giant planet also reaches sufficiently high mass so that both planets are in the low-density gap.

\section{Mass-radius relation and estimating the mass ratio distribution of adjacent pairs}\label{a:q_distribution}
\begin{figure*}
    \centering
    \includegraphics[scale=0.66]{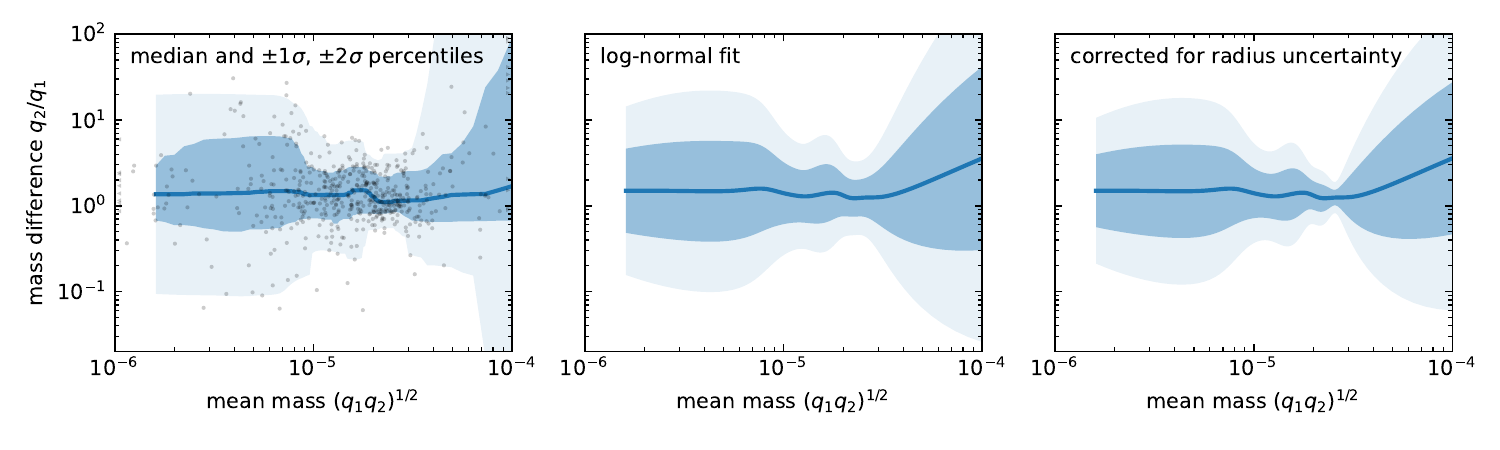}
    \caption{Left panel: Mass ratio of adjacent planet pairs from the CKSM sample (black dots). The distribution around a given $(q_1q_2)^{1/2}$ is estimated using a moving window (Appendix \ref{a:q_distribution}); the line and shaded areas show median and percentiles corresponding to $\pm1\sigma$ and $\pm 2\sigma$. Middle panel: Same as the left panel, but now we fit the distribution of mass ratio to a log-normal distribution. Right panel: Updating the width of the log-normal distribution to correct for the increased variance introduced by the uncertainty in planet radii. This is the distribution shown in the bottom panel of Fig. \ref{fig:superearth_uniformity}.}
    \label{fig:q_distribution}
\end{figure*}
The masses of the CKSM \citep{Weiss2018} transiting exoplanets plotted in Figs. \ref{fig:superearth} and \ref{fig:superearth_uniformity} are estimated using the \texttt{Forecaster} package \citep{ChenKipping2017}, which can estimate a posterior distribution of planet mass from the radius and radius uncertainty of the planet. We take the median of the mass posterior as the estimated mass of the planet. We comment that switching to the simpler mass-radius relation used in \citet{Weiss2018} does not qualitatively affect our results.

The distribution of mass difference in Fig. \ref{fig:superearth_uniformity} is estimated as follows. Consider the sample consisting of all adjacent planet pairs in all systems in the CKSM sample. We sort all pairs by $(q_1q_2)^{1/2}$ and use the sorted index $i$ to label each pair. On this sample, we define a weighting $w_i$ that corresponds to a smooth moving window,
\begin{equation}
    w_i = \exp\left(-\frac{(i-i_{\rm c})^2}{2n^2}\right).
\end{equation}
Here $n$ is the width of this window, and we choose $n=30$; $i_{\rm c}$ is the center of the window which moves across the whole sample. For each $i_{\rm c}$, we use the weighted sample to represent the properties at the $(q_1/q_2)^{1/2}$ value corresponding to the logarithmic average of this weighted sample. Moving $i_{\rm c}$ thus allows us to estimate the distribution of $q_1/q_2$ at different $(q_1/q_2)^{1/2}$.
In Fig. \ref{fig:q_distribution} left panel we plot the median and the percentiles corresponding to $\pm1\sigma$ ($68\%$) and $\pm2\sigma$ ($95\%$) of this distribution. We also plot the result of fitting the $q_1/q_2$ distribution for each $i_{\rm c}$ to a log-normal distribution in the middle panel; comparing the left and middle panels shows that the data is relatively well described by a log-normal distribution for $q\lesssim 5\times 10^{-5}$.

A limitation of the above result is that the uncertainty in planet radius can translate into uncertainties of mass and pairwise mass ratio, which widens the mass ratio distribution. It is easy to correct for the effect of radius uncertainty when we take the assumption that the uncertainties in planet radii $\log r_{1,2}$ and the distribution in $\log q_1/q_2$ are all Gaussian and independent of one another. In that case,
\begin{equation}
    \sigma^2_{\rm obs}(\log q_1/q_2) = \sigma^2(\log q_1/q_2) + \sum_{i=1,2}\left(\frac{\partial\log m_i}{\partial\log r_i}\right)^2\sigma^2(\log r_i).
\end{equation}
Here the $\sigma^2_{\rm obs}(\log q_1/q_2)$ is the estimated variance of $\log q_1/q_2$ based on the observed radii (as plotted in Fig. \ref{fig:q_distribution} middle panel); $\sigma^2(\log q_1/q_2)$ is the variance of $\log q_1/q_2$ based on the true radii; and the last term on the right hand side is the variances due to the uncertainties in planet radii. Following \citet[Fig. 3]{ChenKipping2017}, we take
\begin{equation}
    \frac{\partial\log m}{\partial\log r} = 
    \begin{cases}
        0.28^{-1} & m<2M_\oplus,\\
        0.59^{-1} & 2M_\oplus\leq m<0.4M_{\rm Jup},\\
        -0.04^{-1} & 0.4M_{\rm Jup}\leq m<0.08M_\odot,\\
        0.88^{-1} & m\geq 0.08M_\odot.
    \end{cases}
\end{equation}
This allows us to estimate $\sigma^2(\log q_1/q_2)$ from $\sigma^2_{\rm obs}(\log q_1/q_2)$ and the radius uncertainty $\sigma^2(\log r_i)$, and the resulting distribution is plotted in the right panel of Fig. \ref{fig:q_distribution} and the bottom panel of Fig. \ref{fig:superearth_uniformity}.

Although we have accounted for errors due to radius uncertainty, in this analysis we do not attempt to account for errors due to the deviation from the mass-radius relation we adopted, because is challenging to estimate its impact on $q_1/q_2$. The deviation from a given mass-radius relation can be interpreted as the sum of a inter-system component and an intra-system component; the former is constant within each system and does not affect the mass ratios of intra-system pairs. Although the mass posterior produced by \texttt{Forecaster} includes the uncertainty in the mass-radius relation, it is unclear what fraction of this uncertainty corresponds to intra-system variation.

\section{Statistical significance of heteroscedasticity}\label{a:heteroscedasticity}
The uniformity dichotomy between Earths and Super-Earths discussed in Section \ref{sec:superearths:obs} (Fig. \ref{fig:superearth_uniformity}) can be phrased as a heteroscedasticity between the pairwise mass difference $\log q_1/q_2$ and the mean planet mass $\log (q_1q_2)^{1/2}$. When all planets in the sample form via the same mechanisms, we expect the scatter in $\log q_1/q_2$ to be constant (or homoscedastic); and deviation from that (heteroscedasticity) signifies the existence of some mass-dependent formation mechanism (e.g., pressure bumps). In this appendix we test the statistical significance of heteroscedasticity in our sample using two commonly applied diagnostics, the Breusch-Pagan test and the White test \citep{kaufman2013heteroskedasticity}. Both tests test against the null hypothesis that the data is homoscedastic. We first perform the test on all adjacent pairs in the CKSM sample; this gives highly statistically significant $p$-values, with $p_{\rm W}=6\times 10^{-31}$ for the White test and $p_{\rm BP} = 1\times 10^{-15}$ for the Breusch-Pagan test. To focus on the comparison between Earths and Super-Earths, we restrict our sample to low-mass planets with $(q_1q_2)^{1/2} < 3\times 10^{-5}$ ($\lesssim 10M_\oplus$ for a sun-like star). The result remains highly statistically significant, with $p_{\rm W}=1\times 10^{-11}$ and $p_{\rm BP} = 3\times 10^{-6}$, confirming the visual trend in Fig. \ref{fig:superearth_uniformity}.

\bibliography{XW23}{}

\begin{thebibliography}{}
\expandafter\ifx\csname natexlab\endcsname\relax\def\natexlab#1{#1}\fi
\providecommand{\url}[1]{\href{#1}{#1}}
\providecommand{\dodoi}[1]{doi:~\href{http://doi.org/#1}{\nolinkurl{#1}}}
\providecommand{\doeprint}[1]{\href{http://ascl.net/#1}{\nolinkurl{http://ascl.net/#1}}}
\providecommand{\doarXiv}[1]{\href{https://arxiv.org/abs/#1}{\nolinkurl{https://arxiv.org/abs/#1}}}

\bibitem[{{Adams}(2019)}]{Adams2019}
{Adams}, F.~C. 2019, \mnras, 488, 1446, \dodoi{10.1093/mnras/stz1832}

\bibitem[{{Agol} {et~al.}(2021){Agol}, {Dorn}, {Grimm}, {Turbet}, {Ducrot},
  {Delrez}, {Gillon}, {Demory}, {Burdanov}, {Barkaoui}, {Benkhaldoun},
  {Bolmont}, {Burgasser}, {Carey}, {de Wit}, {Fabrycky}, {Foreman-Mackey},
  {Haldemann}, {Hernandez}, {Ingalls}, {Jehin}, {Langford}, {Leconte},
  {Lederer}, {Luger}, {Malhotra}, {Meadows}, {Morris}, {Pozuelos}, {Queloz},
  {Raymond}, {Selsis}, {Sestovic}, {Triaud}, \& {Van Grootel}}]{Agol2021}
{Agol}, E., {Dorn}, C., {Grimm}, S.~L., {et~al.} 2021, \psj, 2, 1,
  \dodoi{10.3847/PSJ/abd022}

\bibitem[{{Andama} {et~al.}(2022){Andama}, {Ndugu}, {Anguma}, \&
  {Jurua}}]{Andama2022}
{Andama}, G., {Ndugu}, N., {Anguma}, S.~K., \& {Jurua}, E. 2022, \mnras, 510,
  1298, \dodoi{10.1093/mnras/stab3508}

\bibitem[{{Batygin} \& {Morbidelli}(2023)}]{BatyginMorbidelli2023}
{Batygin}, K., \& {Morbidelli}, A. 2023, Nature Astronomy, 7, 330,
  \dodoi{10.1038/s41550-022-01850-5}

\bibitem[{{Bitsch}(2019)}]{Bitsch2019}
{Bitsch}, B. 2019, \aap, 630, A51, \dodoi{10.1051/0004-6361/201935877}

\bibitem[{{Bitsch} {et~al.}(2018){Bitsch}, {Morbidelli}, {Johansen}, {Lega},
  {Lambrechts}, \& {Crida}}]{Bitsch2018}
{Bitsch}, B., {Morbidelli}, A., {Johansen}, A., {et~al.} 2018, \aap, 612, A30,
  \dodoi{10.1051/0004-6361/201731931}

\bibitem[{{Carrera} {et~al.}(2021){Carrera}, {Simon}, {Li}, {Kretke}, \&
  {Klahr}}]{Carrera2021}
{Carrera}, D., {Simon}, J.~B., {Li}, R., {Kretke}, K.~A., \& {Klahr}, H. 2021,
  \aj, 161, 96, \dodoi{10.3847/1538-3881/abd4d9}

\bibitem[{{Chambers}(2021)}]{Chambers2021}
{Chambers}, J. 2021, \apj, 914, 102, \dodoi{10.3847/1538-4357/abfaa4}

\bibitem[{{Chen} \& {Kipping}(2017)}]{ChenKipping2017}
{Chen}, J., \& {Kipping}, D. 2017, \apj, 834, 17,
  \dodoi{10.3847/1538-4357/834/1/17}

\bibitem[{{Cresswell} \& {Nelson}(2008)}]{CresswellNelson2008}
{Cresswell}, P., \& {Nelson}, R.~P. 2008, \aap, 482, 677,
  \dodoi{10.1051/0004-6361:20079178}

\bibitem[{{Dai} {et~al.}(2020){Dai}, {Winn}, {Schlaufman}, {Wang}, {Weiss},
  {Petigura}, {Howard}, \& {Fang}}]{Dai2020}
{Dai}, F., {Winn}, J.~N., {Schlaufman}, K., {et~al.} 2020, \aj, 159, 247,
  \dodoi{10.3847/1538-3881/ab88b8}

\bibitem[{{Deck} {et~al.}(2013){Deck}, {Payne}, \& {Holman}}]{Deck2013}
{Deck}, K.~M., {Payne}, M., \& {Holman}, M.~J. 2013, \apj, 774, 129,
  \dodoi{10.1088/0004-637X/774/2/129}

\bibitem[{{Dong} {et~al.}(2018){Dong}, {Li}, {Chiang}, \& {Li}}]{Dong2018}
{Dong}, R., {Li}, S., {Chiang}, E., \& {Li}, H. 2018, \apj, 866, 110,
  \dodoi{10.3847/1538-4357/aadadd}

\bibitem[{{Dr{\k{a}}{\.z}kowska} \& {Alibert}(2017)}]{Drazkowska2017}
{Dr{\k{a}}{\.z}kowska}, J., \& {Alibert}, Y. 2017, \aap, 608, A92,
  \dodoi{10.1051/0004-6361/201731491}

\bibitem[{{Dr{\k{a}}{\.z}kowska} {et~al.}(2023){Dr{\k{a}}{\.z}kowska},
  {Bitsch}, {Lambrechts}, {Mulders}, {Harsono}, {Vazan}, {Liu}, {Ormel},
  {Kretke}, \& {Morbidelli}}]{Drazkowska2023}
{Dr{\k{a}}{\.z}kowska}, J., {Bitsch}, B., {Lambrechts}, M., {et~al.} 2023, in
  Astronomical Society of the Pacific Conference Series, Vol. 534, Protostars
  and Planets VII, ed. S.~{Inutsuka}, Y.~{Aikawa}, T.~{Muto}, K.~{Tomida}, \&
  M.~{Tamura}, 717, \dodoi{10.48550/arXiv.2203.09759}

\bibitem[{{Dullemond} {et~al.}(2018){Dullemond}, {Birnstiel}, {Huang},
  {Kurtovic}, {Andrews}, {Guzm{\'a}n}, {P{\'e}rez}, {Isella}, {Zhu}, {Benisty},
  {Wilner}, {Bai}, {Carpenter}, {Zhang}, \& {Ricci}}]{dsharpVI}
{Dullemond}, C.~P., {Birnstiel}, T., {Huang}, J., {et~al.} 2018, \apjl, 869,
  L46, \dodoi{10.3847/2041-8213/aaf742}

\bibitem[{{Emsenhuber} {et~al.}(2023){Emsenhuber}, {Mordasini}, \&
  {Burn}}]{Emsenhuber2023}
{Emsenhuber}, A., {Mordasini}, C., \& {Burn}, R. 2023, European Physical
  Journal Plus, 138, 181, \dodoi{10.1140/epjp/s13360-023-03784-x}

\bibitem[{{Ginzburg} \& {Chiang}(2019)}]{GinzburgChiang2019}
{Ginzburg}, S., \& {Chiang}, E. 2019, \mnras, 487, 681,
  \dodoi{10.1093/mnras/stz1322}

\bibitem[{{Gladman}(1993)}]{Gladman1993}
{Gladman}, B. 1993, \icarus, 106, 247, \dodoi{10.1006/icar.1993.1169}

\bibitem[{{Goldreich} {et~al.}(2004){Goldreich}, {Lithwick}, \&
  {Sari}}]{Goldreich2004}
{Goldreich}, P., {Lithwick}, Y., \& {Sari}, R. 2004, \apj, 614, 497,
  \dodoi{10.1086/423612}

\bibitem[{{Goldreich} \& {Schlichting}(2014)}]{Goldreich2014}
{Goldreich}, P., \& {Schlichting}, H.~E. 2014, \aj, 147, 32,
  \dodoi{10.1088/0004-6256/147/2/32}

\bibitem[{{Goyal} \& {Wang}(2022)}]{Goyal2022}
{Goyal}, A.~V., \& {Wang}, S. 2022, \apj, 933, 162,
  \dodoi{10.3847/1538-4357/ac7562}

\bibitem[{{Guilera} {et~al.}(2020){Guilera}, {S{\'a}ndor}, {Ronco},
  {Venturini}, \& {Miller Bertolami}}]{Guilera2020}
{Guilera}, O.~M., {S{\'a}ndor}, Z., {Ronco}, M.~P., {Venturini}, J., \& {Miller
  Bertolami}, M.~M. 2020, \aap, 642, A140, \dodoi{10.1051/0004-6361/202038458}

\bibitem[{Hartigan \& Hartigan(1985)}]{hartigan1985dip}
Hartigan, J.~A., \& Hartigan, P.~M. 1985, The annals of Statistics, 70

\bibitem[{{Helled} \& {Fortney}(2020)}]{Helled2020}
{Helled}, R., \& {Fortney}, J.~J. 2020, Philosophical Transactions of the Royal
  Society of London Series A, 378, 20190474, \dodoi{10.1098/rsta.2019.0474}

\bibitem[{{Huang} {et~al.}(2018){Huang}, {Andrews}, {Dullemond}, {Isella},
  {P{\'e}rez}, {Guzm{\'a}n}, {{\"O}berg}, {Zhu}, {Zhang}, {Bai}, {Benisty},
  {Birnstiel}, {Carpenter}, {Hughes}, {Ricci}, {Weaver}, \&
  {Wilner}}]{dsharpII}
{Huang}, J., {Andrews}, S.~M., {Dullemond}, C.~P., {et~al.} 2018, \apjl, 869,
  L42, \dodoi{10.3847/2041-8213/aaf740}

\bibitem[{{Jiang} \& {Ormel}(2023)}]{Jiang2023}
{Jiang}, H., \& {Ormel}, C.~W. 2023, \mnras, 518, 3877,
  \dodoi{10.1093/mnras/stac3275}

\bibitem[{{Kanagawa} {et~al.}(2018){Kanagawa}, {Tanaka}, \&
  {Szuszkiewicz}}]{Kanagawa2018}
{Kanagawa}, K.~D., {Tanaka}, H., \& {Szuszkiewicz}, E. 2018, \apj, 861, 140,
  \dodoi{10.3847/1538-4357/aac8d9}

\bibitem[{Kaufman(2013)}]{kaufman2013heteroskedasticity}
Kaufman, R.~L. 2013, Heteroskedasticity in regression: Detection and correction
  (Sage Publications)

\bibitem[{{Lambrechts} {et~al.}(2014){Lambrechts}, {Johansen}, \&
  {Morbidelli}}]{Lambrechts2014}
{Lambrechts}, M., {Johansen}, A., \& {Morbidelli}, A. 2014, \aap, 572, A35,
  \dodoi{10.1051/0004-6361/201423814}

\bibitem[{{Lambrechts} {et~al.}(2019){Lambrechts}, {Morbidelli}, {Jacobson},
  {Johansen}, {Bitsch}, {Izidoro}, \& {Raymond}}]{Lambrechts2019}
{Lambrechts}, M., {Morbidelli}, A., {Jacobson}, S.~A., {et~al.} 2019, \aap,
  627, A83, \dodoi{10.1051/0004-6361/201834229}

\bibitem[{{Lau} {et~al.}(2022){Lau}, {Dr{\k{a}}{\.z}kowska}, {Stammler},
  {Birnstiel}, \& {Dullemond}}]{Lau2022}
{Lau}, T. C.~H., {Dr{\k{a}}{\.z}kowska}, J., {Stammler}, S.~M., {Birnstiel},
  T., \& {Dullemond}, C.~P. 2022, \aap, 668, A170,
  \dodoi{10.1051/0004-6361/202244864}

\bibitem[{{Lesur} {et~al.}(2022){Lesur}, {Ercolano}, {Flock}, {Lin}, {Yang},
  {Barranco}, {Benitez-Llambay}, {Goodman}, {Johansen}, {Klahr}, {Laibe},
  {Lyra}, {Marcus}, {Nelson}, {Squire}, {Simon}, {Turner}, {Umurhan}, \&
  {Youdin}}]{Lesur2022}
{Lesur}, G., {Ercolano}, B., {Flock}, M., {et~al.} 2022, arXiv e-prints,
  arXiv:2203.09821, \dodoi{10.48550/arXiv.2203.09821}

\bibitem[{{Lu} {et~al.}(2020){Lu}, {Schlaufman}, \& {Cheng}}]{Lu2020}
{Lu}, C.~X., {Schlaufman}, K.~C., \& {Cheng}, S. 2020, \aj, 160, 253,
  \dodoi{10.3847/1538-3881/abb773}

\bibitem[{{Luque} \& {Pall{\'e}}(2022)}]{Luque2022}
{Luque}, R., \& {Pall{\'e}}, E. 2022, Science, 377, 1211,
  \dodoi{10.1126/science.abl7164}

\bibitem[{{Manara} {et~al.}(2018){Manara}, {Morbidelli}, \&
  {Guillot}}]{Manara2018}
{Manara}, C.~F., {Morbidelli}, A., \& {Guillot}, T. 2018, \aap, 618, L3,
  \dodoi{10.1051/0004-6361/201834076}

\bibitem[{{Millholland} {et~al.}(2017){Millholland}, {Wang}, \&
  {Laughlin}}]{Millholland2017}
{Millholland}, S., {Wang}, S., \& {Laughlin}, G. 2017, \apjl, 849, L33,
  \dodoi{10.3847/2041-8213/aa9714}

\bibitem[{{Mishra} {et~al.}(2021){Mishra}, {Alibert}, {Leleu}, {Emsenhuber},
  {Mordasini}, {Burn}, {Udry}, \& {Benz}}]{Mishra2021}
{Mishra}, L., {Alibert}, Y., {Leleu}, A., {et~al.} 2021, \aap, 656, A74,
  \dodoi{10.1051/0004-6361/202140761}

\bibitem[{{Mishra} {et~al.}(2023){Mishra}, {Alibert}, {Udry}, \&
  {Mordasini}}]{Mishra2023}
{Mishra}, L., {Alibert}, Y., {Udry}, S., \& {Mordasini}, C. 2023, \aap, 670,
  A68, \dodoi{10.1051/0004-6361/202243751}

\bibitem[{{Morbidelli}(2020)}]{Morbidelli2020}
{Morbidelli}, A. 2020, \aap, 638, A1, \dodoi{10.1051/0004-6361/202037983}

\bibitem[{{Mulders} {et~al.}(2021){Mulders}, {Pascucci}, {Ciesla}, \&
  {Fernandes}}]{Mulders2021}
{Mulders}, G.~D., {Pascucci}, I., {Ciesla}, F.~J., \& {Fernandes}, R.~B. 2021,
  \apj, 920, 66, \dodoi{10.3847/1538-4357/ac178e}

\bibitem[{{Najita} \& {Kenyon}(2014)}]{Najita2014}
{Najita}, J.~R., \& {Kenyon}, S.~J. 2014, \mnras, 445, 3315,
  \dodoi{10.1093/mnras/stu1994}

\bibitem[{{NASA Exoplanet Archive}(2023)}]{exoarchive}
{NASA Exoplanet Archive}. 2023, Planetary Systems, Version: 2023-11-17 18:47,
  NExScI-Caltech/IPAC, \dodoi{10.26133/NEA12}

\bibitem[{{Pichierri} {et~al.}(2023){Pichierri}, {Bitsch}, \&
  {Lega}}]{Pichierri2023}
{Pichierri}, G., {Bitsch}, B., \& {Lega}, E. 2023, \aap, 670, A148,
  \dodoi{10.1051/0004-6361/202245196}

\bibitem[{{Pinilla} {et~al.}(2012){Pinilla}, {Benisty}, \&
  {Birnstiel}}]{Pinilla2012}
{Pinilla}, P., {Benisty}, M., \& {Birnstiel}, T. 2012, \aap, 545, A81,
  \dodoi{10.1051/0004-6361/201219315}

\bibitem[{{Pu} \& {Wu}(2015)}]{PuWu2015}
{Pu}, B., \& {Wu}, Y. 2015, \apj, 807, 44, \dodoi{10.1088/0004-637X/807/1/44}

\bibitem[{{Rice} {et~al.}(2006){Rice}, {Armitage}, {Wood}, \&
  {Lodato}}]{Rice2006}
{Rice}, W.~K.~M., {Armitage}, P.~J., {Wood}, K., \& {Lodato}, G. 2006, \mnras,
  373, 1619, \dodoi{10.1111/j.1365-2966.2006.11113.x}

\bibitem[{{Rosenthal} {et~al.}(2021){Rosenthal}, {Fulton}, {Hirsch},
  {Isaacson}, {Howard}, {Dedrick}, {Sherstyuk}, {Blunt}, {Petigura}, {Knutson},
  {Behmard}, {Chontos}, {Crepp}, {Crossfield}, {Dalba}, {Fischer}, {Henry},
  {Kane}, {Kosiarek}, {Marcy}, {Rubenzahl}, {Weiss}, \&
  {Wright}}]{Rosenthal2021}
{Rosenthal}, L.~J., {Fulton}, B.~J., {Hirsch}, L.~A., {et~al.} 2021, \apjs,
  255, 8, \dodoi{10.3847/1538-4365/abe23c}

\bibitem[{{Rosenthal} \& {Murray-Clay}(2020)}]{RosenthalMurrayClay2020}
{Rosenthal}, M.~M., \& {Murray-Clay}, R.~A. 2020, \apj, 898, 108,
  \dodoi{10.3847/1538-4357/ab9eb2}

\bibitem[{{Tanaka} \& {Ward}(2004)}]{TanakaWard2004}
{Tanaka}, H., \& {Ward}, W.~R. 2004, \apj, 602, 388, \dodoi{10.1086/380992}

\bibitem[{{Teague} {et~al.}(2018){Teague}, {Bae}, {Birnstiel}, \&
  {Bergin}}]{Teague2018}
{Teague}, R., {Bae}, J., {Birnstiel}, T., \& {Bergin}, E.~A. 2018, \apj, 868,
  113, \dodoi{10.3847/1538-4357/aae836}

\bibitem[{{Teanby} {et~al.}(2020){Teanby}, {Irwin}, {Moses}, \&
  {Helled}}]{Teanby2020}
{Teanby}, N.~A., {Irwin}, P.~G.~J., {Moses}, J.~I., \& {Helled}, R. 2020,
  Philosophical Transactions of the Royal Society of London Series A, 378,
  20190489, \dodoi{10.1098/rsta.2019.0489}

\bibitem[{{Wang}(2017)}]{Wang2017}
{Wang}, S. 2017, Research Notes of the American Astronomical Society, 1, 26,
  \dodoi{10.3847/2515-5172/aa9be5}

\bibitem[{{Weiss} {et~al.}(2022){Weiss}, {Millholland}, {Petigura}, {Adams},
  {Batygin}, {Bloch}, \& {Mordasini}}]{Weiss2022}
{Weiss}, L.~M., {Millholland}, S.~C., {Petigura}, E.~A., {et~al.} 2022, arXiv
  e-prints, arXiv:2203.10076, \dodoi{10.48550/arXiv.2203.10076}

\bibitem[{{Weiss} {et~al.}(2018){Weiss}, {Marcy}, {Petigura}, {Fulton},
  {Howard}, {Winn}, {Isaacson}, {Morton}, {Hirsch}, {Sinukoff}, {Cumming},
  {Hebb}, \& {Cargile}}]{Weiss2018}
{Weiss}, L.~M., {Marcy}, G.~W., {Petigura}, E.~A., {et~al.} 2018, \aj, 155, 48,
  \dodoi{10.3847/1538-3881/aa9ff6}

\bibitem[{{Wu} {et~al.}(2023){Wu}, {Rice}, \& {Wang}}]{Wu2023}
{Wu}, D.-H., {Rice}, M., \& {Wang}, S. 2023, \aj, 165, 171,
  \dodoi{10.3847/1538-3881/acbf3f}

\bibitem[{{Xu}(2022)}]{Xu2022}
{Xu}, W. 2022, \apj, 934, 156, \dodoi{10.3847/1538-4357/ac7b94}

\bibitem[{{Xu} \& {Armitage}(2023)}]{XuArmitage2023}
{Xu}, W., \& {Armitage}, P.~J. 2023, \apj, 946, 94,
  \dodoi{10.3847/1538-4357/acb7e5}

\bibitem[{{Xu} {et~al.}(2023){Xu}, {Ohashi}, {Aso}, \& {Liu}}]{Xu2023}
{Xu}, W., {Ohashi}, S., {Aso}, Y., \& {Liu}, H.~B. 2023, arXiv e-prints,
  arXiv:2308.01972, \dodoi{10.48550/arXiv.2308.01972}

\bibitem[{{Xu} \& {Bai}(2022)}]{XuBai2022}
{Xu}, Z., \& {Bai}, X.-N. 2022, \apjl, 937, L4,
  \dodoi{10.3847/2041-8213/ac8dff}

\bibitem[{{Zhang} {et~al.}(2018){Zhang}, {Zhu}, {Huang}, {Guzm{\'a}n},
  {Andrews}, {Birnstiel}, {Dullemond}, {Carpenter}, {Isella}, {P{\'e}rez},
  {Benisty}, {Wilner}, {Baruteau}, {Bai}, \& {Ricci}}]{dsharpVII}
{Zhang}, S., {Zhu}, Z., {Huang}, J., {et~al.} 2018, \apjl, 869, L47,
  \dodoi{10.3847/2041-8213/aaf744}

\end{thebibliography}
\bibliographystyle{aasjournal}

\end{document}